\documentclass[twoside]{IEEEtran}
\usepackage{atbegshi,picture,ifpdf,setspace}
\usepackage[noadjust]{cite}
\usepackage[pdftex]{graphicx}
\usepackage[cmex10]{amsmath}
\usepackage[caption=false,font=footnotesize]{subfig}
\usepackage{url}

\usepackage[english]{babel}
\usepackage[utf8x]{inputenc}
\usepackage[T1]{fontenc}
\usepackage[cmintegrals,varvw]{newtxmath}
\usepackage[mathscr]{euscript}
\usepackage{bm}
\usepackage{siunitx,xcolor,ellipsis}

\makeatletter
\def\ps@IEEEtitlepagestyle{%
  \def\@oddfoot{\mycopyrightnotice}%
  \def\@oddhead{\hbox{}\@IEEEheaderstyle\leftmark\hfil\thepage}\relax
  \def\@evenhead{\@IEEEheaderstyle\thepage\hfil\leftmark\hbox{}}\relax
  \def\@evenfoot{}%
}
\def\mycopyrightnotice{%
  \begin{minipage}{\textwidth}
  \centering \scriptsize
  Copyright~\copyright~2017 IEEE. Personal use of this material is permitted. Permission from IEEE must be obtained for all other uses, in any current or future media, including\\reprinting/republishing this material for advertising or promotional purposes, creating new collective works, for resale or redistribution to servers or lists, or reuse of any copyrighted component of this work in other works by sending a request to pubs-permissions@ieee.org.
  \end{minipage}
}
\makeatother

\AtBeginShipout{\AtBeginShipoutUpperLeft{%
  \put(\dimexpr\paperwidth-1cm\relax,-.55cm){\makebox[0pt][r]{
  \begin{minipage}{\textwidth}
  \centering \scriptsize 
  This is the author's version of an article that has been published in this journal. 
  Changes were made to this version by the publisher prior to publication. 
  \\[0.3ex]
  The final version of record is available at\qquad http://dx.doi.org/10.1109/TAP.2017.2780897
  \end{minipage}
  }}}%
}

\begin{document}
\title{A Millimeter-Wave Self-Mixing Array with Large Gain and Wide Angular Receiving Range}
\author{Jonas Kornprobst, \IEEEmembership{Student Member, IEEE}, Thomas J. Mittermaier, \IEEEmembership{Member, IEEE}, and\\Thomas F. Eibert, \IEEEmembership{Senior Member, IEEE}%
\thanks{Manuscript received March 14, 2017; revised October 19, 2017; accepted November 19, 2017. This work was supported by the Deutsche Forschungsgemeinschaft (DFG) under Grant EI 352/17-1. \emph{(Corresponding author: Jonas Kornprobst.)}}%
\thanks{J. Kornprobst and T. F. Eibert are with the Chair of High-Frequency Engineering, Technical University  of  Munich,  80290  Munich, Germany (e-mail:  j.kornprobst@tum.de;hft@ei.tum.de).}%
\thanks{T. J. Mittermaier was with the Chair of High-Frequency Engineering, Technical University  of  Munich,  80290  Munich, Germany  and is now with RF360 Europe GmbH, Munich, Germany.}%
\thanks{Color versions of one or more of the figures in this communication are available online at http://ieeexplore.ieee.org}
\thanks{Digital  Object Identifier  10.1109/TAP.2017.2780897}
}
\markboth{Transactions on Antennas and Propagation}%
{Kornprobst \MakeLowercase{\textit{et al.}}: A Millimeter-Wave Self-Mixing Array with Large Gain and Wide Angular Receiving Range}

\maketitle

\begin{abstract}
The concept of self-mixing antenna arrays is presented and analyzed with respect to its beneficial behavior of large gain over a wide angular range. 
The large gain is attained by an antenna array with large element spacing, where all array element signals are combined approximately coherently over the entire angular receiving range. 
This functionality is achieved by the self-mixing principle, where an exact description via an intermediate frequency (IF) array factor is derived. 
For verification purposes, a 4$\,\boldsymbol\times\,$2 self-mixing array is fabricated and measured in the frequency range from 34\,GHz to 39\,GHz. 
A multiple-resonances millimeter-wave microstrip patch antenna has been especially developed to achieve large bandwidth and a wide angular receiving range. 
The broad beamwidth is achieved by two parasitic patches and suitable radiation characteristics of the resonant modes. 
The self-mixing of the receive signal is realized at each antenna element by a Schottky diode with an optimized operating point. 
The down-converted array element signals are then combined and measured at the IF. 
The receive power is increased significantly over a large angular range as compared to conventional array feeding techniques. The simulation results are verified by measurements, which show very good agreement.  
\end{abstract}


\section{Introduction} 
\IEEEPARstart{F}{or the} increasing demand of ever-expanding data rates in wireless communications, millimeter-wave (mm-wave) frequencies are widely investigated, e.g. for the fifth generation mobile communication standard~\cite{Rappaport13,Andrews2014,Agiwal16} or the Wifi standard WiGig~\cite{Perahia10,Hansen11}. 
Data links at such high radio frequencies (RF) enable higher data rates, corresponding to an increased channel capacity obtained by the utilization of wide-band radio channels. 
Yet, mm-wave communication suffers significantly from the high path loss according to the Friis transmission equation~\cite{Friis}. 
Therefore, antennas or antenna arrays should have a large gain to compensate for the path loss~\cite{Roh14,Hur13,Wei14,Niu2015}. 
For commonly employed passive and reciprocal antennas, the gain $G=4\uppi A_{\mathrm{eff}}/\lambda^{2}$ is directly related to the effective area~$A_{\mathrm{eff}}$ of the antenna and the free-space wavelength~$\lambda$~\cite{Balanis2016}. 
Furthermore, from the relationship between gain and directivity~$G=\eta D$, where~$\eta$ is the antenna efficiency, it is obvious that reciprocal large gain antennas and antenna arrays are necessarily highly directive.  
Therefore, typically studied antenna arrays for mobile applications in the mm-wave regime are very directive~\cite{Sun13,Li14,Li14b,Semkin15,Khalily16,Ojaroudiparchin16}. 
The high directivity can be of disadvantage, since beamforming is necessary in a very exact manner to maintain a wireless link. 
It can be quite tedious to align the antenna beams exactly, especially if the antennas are integrated in mobile devices. 
Thus, a very advantageous scenario could consist of a base station, which is able to perform beamforming with a sharp beam, but a user with a wide-beam receive pattern, of course also with large gain but with no need for beamforming. 
It is clear that this requirement violates reciprocity. 

In the presented self-mixing concept, the frequency conversion to the lower frequency band is performed without local oscillator (LO) for the receive case only. Separate techniques are still necessary for the transmit scenario. 
Recent research indicates that large gain antennas with wide beam are possible via self-mixing arrays in the receive case with a square-law detector at each antenna, in contrast to common super-heterodyne receivers  with conventional beamforming~\cite{Choi_2007,Wang15}. 
Self-mixing arrays can be deployed in conjunction with the self-heterodyne transmission technique, where a carrier is transmitted in addition to e.g. an orthogonal frequency division multiplexing (OFDM) signal~\cite{Shoji_2001,Shoji_2002,Shoji_2004}. 
Thereby, the necessity of an LO is avoided, which has a positive influence regarding phase noise and frequency offset correction of the receive signal, what is especially important for high frequencies in the mm-wave regime~\cite{Nakatogawa_2006,CHOI_2010,Pacheco}. 
In particular for OFDM signals, radio frequency (RF) phase noise is a crucial point~\cite{Bladel,Armada1998}. 
Therefore, the self-heterodyne technique is a promising approach at very high frequencies~\cite{Foulon2014,Song16}, but also in other research fields, e.g. in optics~\cite{Yin10,Lin_2011,Kanno11,Shao12,tan201464qam,Bergman16,Duarte16} or acoustic underwater communications~\cite{Wiedmann12,Wiedmann2013,Wiedmann16}. 
Recent research in the field of self-heterodyne transmission techniques concentrates on studying the effects of self-mixing arrays~\cite{Wang15} and on improving the transmission technique on a signal processing level~\cite{Choi11,Fernando_2013,Fernando_2013b,Fernando_2013c,Fernando16} or even on extending the self-mixing technique to a more general approach without an explicitly transmitted carrier signal~\cite{ars2017}. 
Also, the analogue receiver circuit can be improved as compared to a simple square-law detector by employing the self-coherent technique~\cite{Jin16}. 

In this paper, a self-mixing antenna array is presented, which achieves constructive superposition of the array element receive signals over a wide angular range.  
A large gain and a large effective area are realized without increasing the directivity significantly. 
A specially designed wide-beamwidth microstrip patch antenna, a low noise amplifier (LNA), and a Schottky diode for down-conversion have been integrated to a self-mixing receive antenna element. 
The receive signals from all the array elements are then combined coherently and a broad receive pattern is obtained, without the need for further signal processing.  

The paper is structured as follows. 
First, the functional principle of the self-mixing antenna array is discussed. 
Afterwards, the antenna and receiver circuit designs are presented. 
Finally, measurements of a single self-mixing antenna element and of a $4\times 2$ self-mixing array are performed to evaluate the array performance and show the excellent agreement with the theoretical considerations.

\section{Basic Principle of the Self-Mixing Array} 
The term \emph{self-mixing receiver} describes a receiver without LO. 
Instead, down-conversion of the receive (Rx) signal is performed by multiplying the time-domain signal by itself. 
Such a square-law detector can easily be realized by a Schottky diode with an ideally quadratic voltage-current  ($v$-$i$) characteristic. 
The  time-domain output signal of such a mixer is given as
\begin{equation}
y(t) = x^{2}(t)
\end{equation} 
with the input signal $x(t)$. 
For a mono-frequent signal, only a direct current (DC) part and a signal with twice the original frequency are obtained. 

\begin{figure}[t]
 \centering
 \includegraphics[]{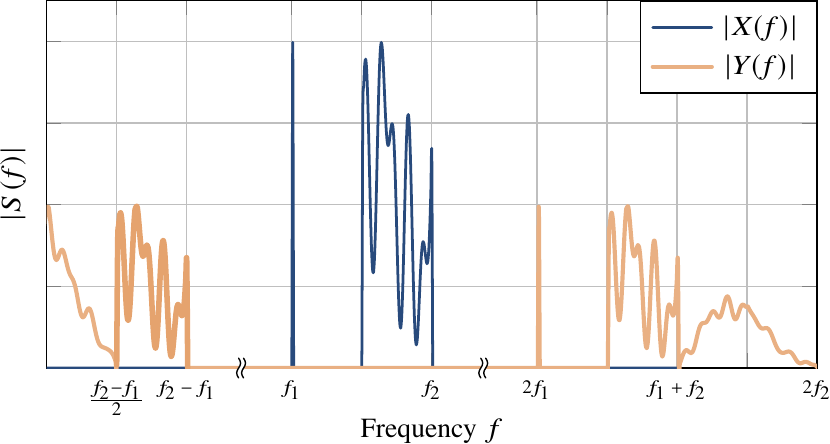}
 \caption{Original spectrum $X(f)$ and self-mixed (down- and up-converted) spectrum $Y(f)$.\label{fig:spectrum}}%
\end{figure}

Any information-carrying signal will necessarily possess a certain RF bandwidth, leading to some of the Rx power spread around DC after self-mixing. 
By transmitting an RF information carrying signal including a separate RF carrier signal, as it is the case for the self-heterodyne transmission technique, the down-converted signal will be found at an intermediate frequency (IF). 
This mixing behavior in the spectrum is illustrated in Fig.~\ref{fig:spectrum}.  
The original spectrum $X(f) = \mathscr{F}\{x(t)\}$ is converted to the Rx spectrum
\begin{equation}
Y(f) = \mathscr F\left\{ x^{2}(t)\right\}=X(f)\ast X(f),
\end{equation}
where  the $\ast$ represents the convolution operator.  
For practical reasons, only the signal parts at lower frequencies are considered for further processing, i.e. the self-mixed signal is low-pass or band-pass filtered.  
In the self-heterodyne  case as depicted in Fig.~\ref{fig:spectrum}, the original information carrying signal located in the range $({f_{1}+f_{2}})/{2}\le f\le f_{2}$ is down-converted into the frequency range $({f_{2}-f_{1}})/{2}\le f\le f_{2}-f_{1}$. 
Therefore, the antenna and mixer analysis will be perfomed with a simplified transmit signal, namely a two-tone signal. 
A very similar conversion to IF is observed and the Rx characteristic can be analyzed in a more convenient way. 
 
\subsection{Receive Characteristic of  a Single Self-Mixing Antenna} %
To analyze the self-mixed signal without loss of generality, a two-tone signal of the form
\begin{equation}
x(t) = \hat{x}_{\mathrm{I}} \sin \omega_{\mathrm{I}} t + \hat{x}_\mathrm{II} \sin\omega_{\mathrm{II}}t 
\end{equation}
is assumed to be transmitted, with the amplitudes $\hat{x}_{\mathrm{I}}$ and $\hat{x}_{\mathrm{II}}$ at the frequencies $f_{\mathrm{I}}=\omega_{\mathrm{I}}/2\uppi$ and $f_{\mathrm{II}}=\omega_{\mathrm{II}}/2\uppi$, respectively. 
The self-mixed signal is attained as
\begin{equation}
y(t) =\frac{1}{2}\left[\hat{x}_{\mathrm{I}}^2 + \hat{x}_{\mathrm{II}}^2 + \hat{x}_{\mathrm{I}}\hat{x}_{\mathrm{II}} \cos((\omega_{\mathrm{I}}-\omega_{\mathrm{II}})t)\right],\label{eq:singleselfmix} 
\end{equation}
if the parts at twice the frequency are dropped after low-pass filtering. 
It becomes obvious from~\eqref{eq:singleselfmix} that the self-mixed IF signal consists of the multiplication of the two received amplitudes. 
In a real-world implementation, these amplitudes are influenced by many frequency-dependent factors, such as transmit power, transmit antenna gain, receive antenna gain, amplifier gain, mixer conversion loss, amplitude and phase LO fluctuations and frequency-dependent path loss. 
However, for a receive pattern measurement, the predominant effect will be the angle-dependent receive gain. 
Therefore, it can be concluded that the receive pattern of a self-mixing antenna
\begin{equation}
C_{\mathrm{sm}}(\vartheta,\varphi) \propto C_{\mathrm{I}}(\vartheta,\varphi)C_{\mathrm{II}}(\vartheta,\varphi)\label{eq:pattern}
\end{equation} 
is in a good approximation the multiplication of the radiation characteristics $C_{{\mathrm{I}},{\mathrm{II}}}$ at the frequencies $f_{{\mathrm{I}},{\mathrm{II}}}$ for an ideal square-law detector, whereby the receive signal is observed at the difference frequency $f_{\mathrm{sm}}=\left|f_{\mathrm{I}}-f_{\mathrm{II}}\right|$. 

\subsection{Receive Characteristic of  a Self-Mixing Antenna Array} %
\begin{figure}[t]
\centering
    \includegraphics[]{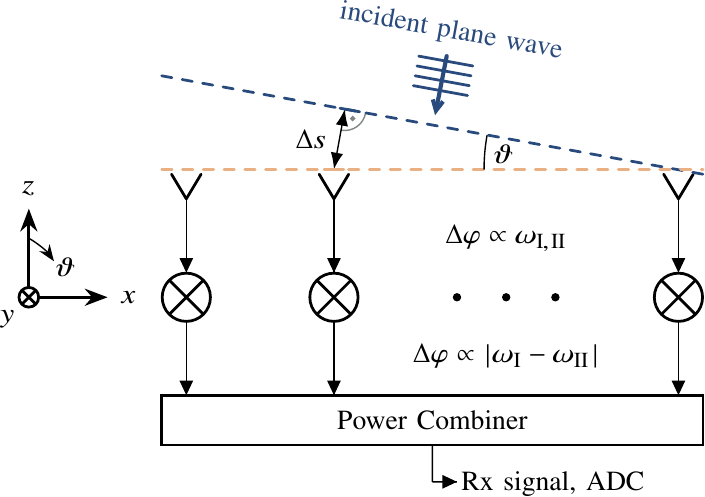}
 \caption{Origin of the self-mixing array factor.\label{fig:array}}
\end{figure}
A schematic, simplified version of a self-mixing antenna array is shown in Fig. \ref{fig:array}. 
Each antenna  element is followed by its own mixer and the signals are combined coherently with equal weighting and equal phase after down-conversion. 
Therefore, the differences in the self-mixed signals, coming from the different antenna elements, have to be analyzed first. 
In principle,  a phase shift
\begin{equation}
\Delta \varphi_{1,2} = 2\uppi\frac{\Delta s}{\lambda_{\mathrm{I},\mathrm{II}}} =2\uppi\frac{\Delta s f_{\mathrm{I},\mathrm{II}}}{c_0}\label{eq:phase}
\end{equation}
is observed between two array elements for plane wave illumination due to a path difference $\Delta s$ of the incident wave, where $c_0$ is the speed of light in free space. 
For a linear array with broadside direction in $+z$-direction, \eqref{eq:phase} can be further modified for the phase shift between the first and the $k$th array element
\begin{equation}
\Delta \varphi_{1,k} = 2\uppi{d_{1,k}}\frac{f_{\mathrm{I},\mathrm{II}}}{c_0}\sin\vartheta
\end{equation}
 with the element distance $d_{1,k}$ in the \textit{xy}-plane. 
The down-converted, band-pass filtered signal at the $k$th antenna reads
\begin{equation}
y_k(t) =\frac{1}{2}\hat{x}_{\mathrm{I}} \hat{x}_{\mathrm{II}}
\cos\left(
(\omega_{\mathrm{I}}-\omega_{\mathrm{II}})t+
{d_{1,k}}\left(\frac{\omega_{{\mathrm{I}}}}{c_0} - \frac{\omega_{\mathrm{II}}}{c_0}\right)\sin\vartheta
\right).
\end{equation}
The most important implication of this result is that the IF phase difference of the self-mixed signal
\begin{equation}
\Delta \varphi_{1,k,\mathrm{sm}} = {d_{1,k}}\frac{\omega_{{\mathrm{I}}}-\omega_{{\mathrm{II}}}}{c_0} \sin\vartheta 
\end{equation}
only depends on the difference between the transmit frequencies. 
Hence, the effective spacing between the antenna elements is reduced and a wider beam can be achieved, even with a sparse placement of the array elements. 
This is very beneficial for the antenna design, since even large antenna elements, occupying more area than e.g.\ $\lambda/2$ by $\lambda/2$, can be employed and, at the same time, it is possible to reduce mutual coupling between the antenna elements to a minimum by a large array spacing. 
The array receive pattern
\begin{equation}
C_{\mathrm{tot}}(\vartheta,\varphi)=C_{\mathrm{array}}(\vartheta,\varphi)C_{\mathrm{sm}}(\vartheta,\varphi)
\end{equation}
is derived from the normalized IF array factor for broadside radiation of an array in the \textit{xy}-plane 
\begin{equation}
C_{\mathrm{array}}(\vartheta,\varphi)=\frac{1}{N}\left|\sum_{k=1}^{N}\mathrm e ^{\,\mathrm{j}d_{1,k} \sin\vartheta({\omega_{\mathrm{I}} - \omega_{\mathrm{II}}})/{c_0}}\right|\label{eq:AF}
\end{equation}
and the multiplication of the antenna radiation patterns at both considered frequencies according to~\eqref{eq:pattern}. 
As compared to an RF array factor of an identical array
\begin{equation}
C_{\mathrm{array,RF}}(\vartheta,\varphi)=\frac{1}{N}\left|\sum_{k=1}^{N}\mathrm e ^{\,\mathrm{ j}d_{1,k} \sin\vartheta(\omega_{\textrm{RF}})/{c_0}}\right|,
\end{equation}
it becomes obvious that such an RF array with wide element spacing will suffer from grating lobes and will have a much smaller beamwidth in $+z$-direction due to the faster phase change in direct proportionality to $\omega_{\mathrm{RF}}\sin\vartheta$. 
In contrast, the phase variation over the incident angle only depends on the much smaller difference frequency $({\omega_{\mathrm{I}}-\omega_{\mathrm{II}}})$ according to~\eqref{eq:AF} in case of self-mixing, where it is clear that self-mixing offers the most benefit if the bandwidth of the received signal is considerably smaller than the RF.

\begin{figure}[t]
\centering
    \includegraphics[]{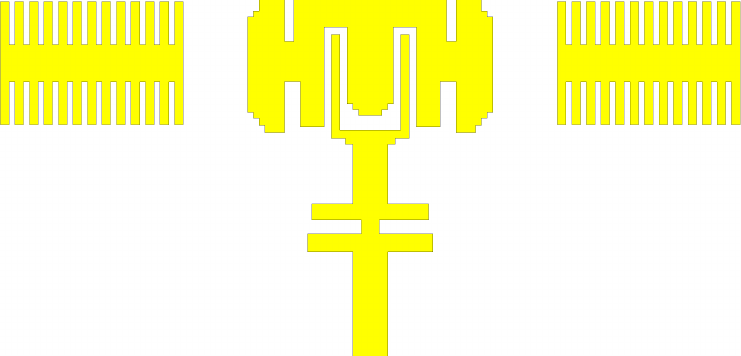}%
 \caption{Antenna structure in top copper layer~\cite{mmwave-ant}.\label{fig:Ant}}
\end{figure}%

Furthermore, in the limiting case $\omega_{\mathrm{I}}-\omega_{\mathrm{II}}\rightarrow0$, the observed phase shift of the array element signals approaches zero according to~\eqref{eq:AF}. 
For~$N$ array elements, this leads to an array gain of~$N$, independent of the angle of incidence.

\section{The Self-Mixing Receiver} 
In this Section, the antenna under test (AUT) Rx module is presented, which consists of a wide-beam broadband patch antenna and a following receive circuit. 
For the down-conversion by self-mixing,  an LNA and a Schottky diode are employed. 
Simulations were performed with Computer Simulation Technology Microwave Studio (CST MWS)~\cite{CST} and Keysight Advanced Design System (ADS)~\cite{ADS}. 

\subsection{Antenna Design} %
\begin{figure}[t]
\centering
    \includegraphics[]{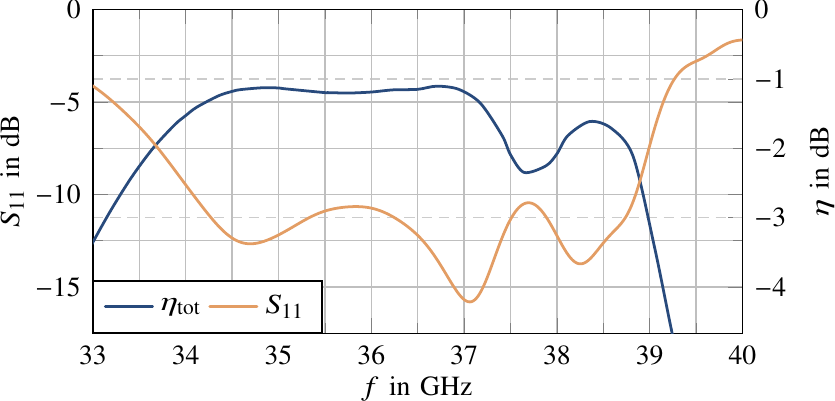}%
 \caption{Simulated $\left|S_{11}\right|$ and $\eta$ of the employed microstrip patch antenna.\label{fig:AUT_S11}}
\end{figure}
The utilized antenna element is a sophisticated design on one-layer Rogers RO3003 substrate~\cite{RO} with 0.25\,mm height with the important properties of broad bandwidth and wide angular range~\cite{mmwave-ant}. 
In Fig.~\ref{fig:Ant}, the copper structure is shown as it is simulated in CST MWS. 
The relative bandwidth achieves more than 13.1\si{\percent} as shown in Fig.~\ref{fig:AUT_S11}, while the total efficiency ranges between $-1$\,dB and $-2$\,dB within the matched bandwidth. 
This broad bandwidth is achieved by several design elements. 
First, the microstrip patch antenna shows two resonances, which is a common \mbox{approach~\cite{yang2001,Lee12,Khan16}.} 
One of the resonant modes is a typical rectangular patch mode, which shows a good impedance bandwidth due to the large patch width~\cite{Jackson1991}. 
Furthermore, the coupling gap feed enhances the bandwidth~\cite{Bhartia}. 
Finally, an additional matching network in microstrip technology was designed to enhance the matched bandwidth~\cite{Pues}. 

\begin{figure}[t]
 \centering
 \includegraphics[]{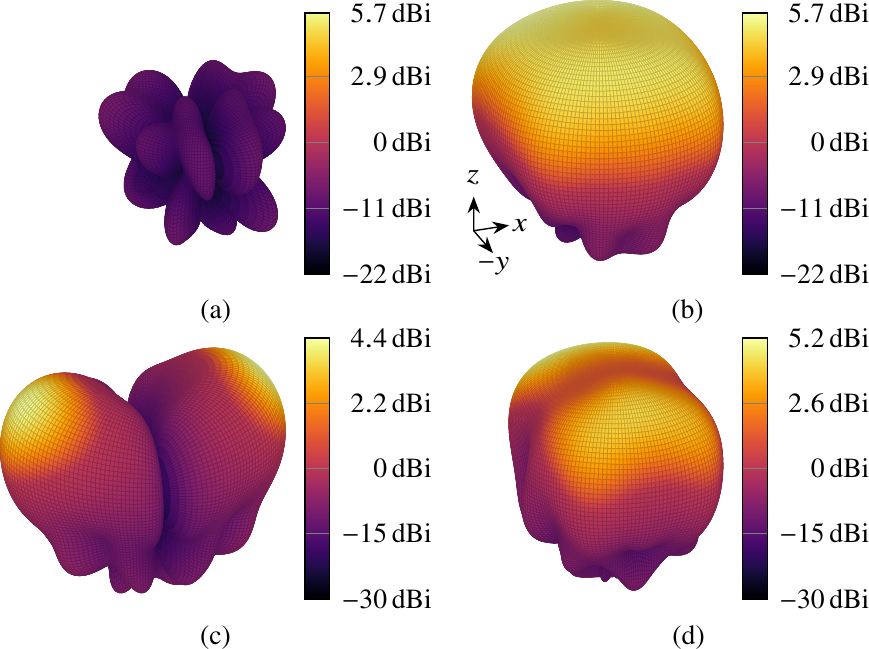}%
 \caption{Radiation patterns of the wide-beam microstrip patch antenna including polarization information. (a)~Ludwig 3 horizontal,  $34.25$\,GHz. (b) Ludwig 3 vertical,  $34.25$\,GHz. (c)~Ludwig 3 horizontal,  $38.25$\,GHz. (d)~Ludwig 3 vertical,  $38.25$\,GHz.\label{fig:patterns}}
\end{figure}

The property of a wide angular receiving range, which is needed for the application in the self-mixing array, is identical to a low directivity and is achieved in two ways. 
The radiation characteristic in the lower frequency range, for the rectangular patch resonant mode, is broadened by the radiation of the two parasitic patches. 
These are designed to be excited out of phase and with lower amplitude to produce destructive interference in $+$\textit{z}-direction. 
For the upper frequency range, the low directivity is attained by the radiation of the second resonant mode, which shows inherently destructive interference in \mbox{$+$\textit{z}-direction} and has, thus, two main beams tilted towards the $\pm x$-axis.
All in all, the directivity is reduced from about $8.5$\,dBi for a common rectangular patch antenna to only $5.5$\,dBi to $7$\,dBi. 
To illustrate the behavior, two exemplary radiation characteristics are given in Fig.~\ref{fig:patterns}. 

\begin{figure}[t]
\centering
    \includegraphics[]{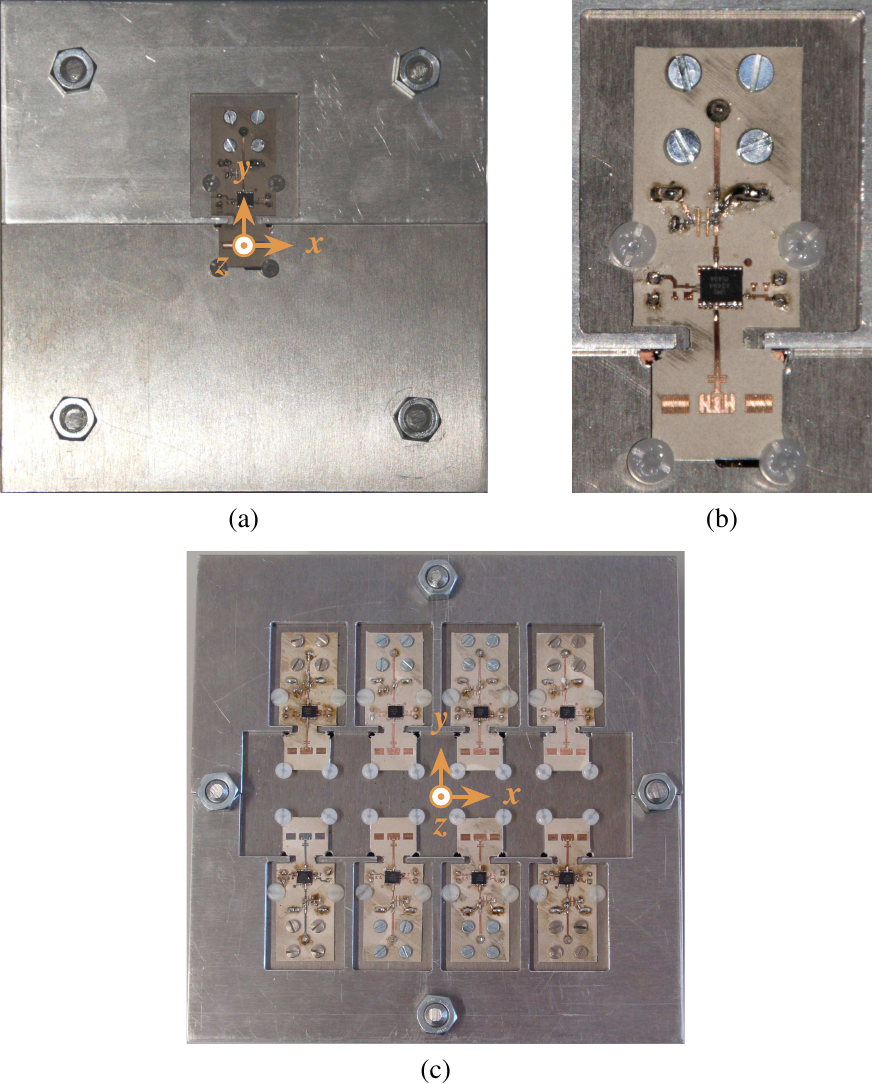}%
 \caption{Photographs of the self-mixing AUTs, with mounting and Rx circuitry. (a)~Single self-mixing AUT on mounting plate. (b)~ Close-up of the single AUT including LNA and mixer. (c)~Self-mixing AUT array on mounting plate.\label{fig:AUT_circuit}}
\end{figure}

The AUT including the Rx circuit is shown in Fig.~\ref{fig:AUT_circuit}. 
Furthermore, the antenna array is shown in Fig.~\ref{fig:AUT_circuit}(c). 
The slotted antenna with parasitic patches and matching network can be recognized as well as the following amplifier and mixer.

\subsection{Receiving Circuit} %
\begin{figure}[t]
 \centering
    \includegraphics[]{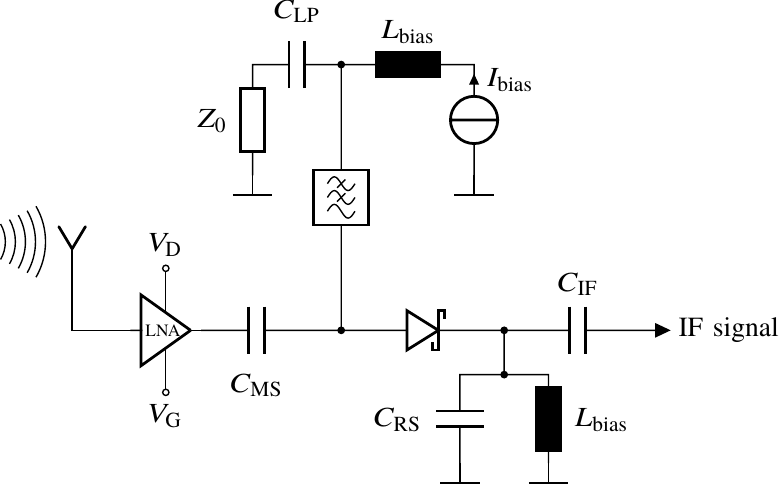}%
 \caption{Schematic of the receiver circuit.\label{fig:AUT_circuit_schem}}
 \end{figure}%

The antenna is directly connected to  a circuit for  self-mixing  the receive signal. 
Fig.~\ref{fig:AUT_circuit_schem} gives the schematic of the receiver circuit including LNA, filters, mixer and supply circuit. 
In general, a  Schottky diode is the component of choice for high-speed mixing stages due to the beneficial behavior of the Schottky barrier at RF~\cite{MaasMixer:1993,Sze:2001}. 
Additionally, they serve excellently as square-law detectors. 

Since the observed IF signal containing the second order mixing products depends quadratically on the input power, i.e. with \SI{3}{dB} more input power \SI{6}{dB} more IF power is attained, it makes sense to place an amplifier just before the mixer. 
Therefore, the LNA CHA2494-QEG~\cite{UMS} with a drain voltage of $V_\mathrm{D}=4\,\mathrm{V}$ and a gate voltage of $V_\mathrm{G}=-0.24\,\mathrm{V}$, to obtain a drain current of $I_\mathrm{D}=\SI{120}{\milli\ampere}$ and, thus, a gain of about~25\,dB in the considered frequency range, is utilized. 
Then, a coupling capacitance~$C_\mathrm{MS}$ in microstrip technology with a gap of 25\,\si{\micro\meter} is placed, exhibiting an insertion loss below 0.25\,dB in the frequency range from 34\,GHz to 39\,GHz.

\begin{figure}[t]
 \centering
 \includegraphics[]{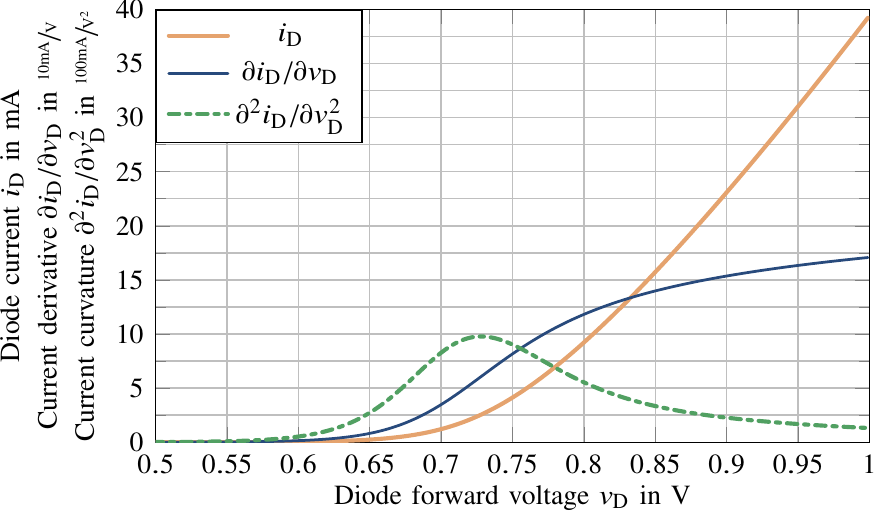}%
 \caption{Analysis of the DC operating point $v$-$i$-characteristic of the employed diode.\label{fig:diode_dc}}
 \end{figure}%

\begin{figure}[t]
 \centering
    \includegraphics[]{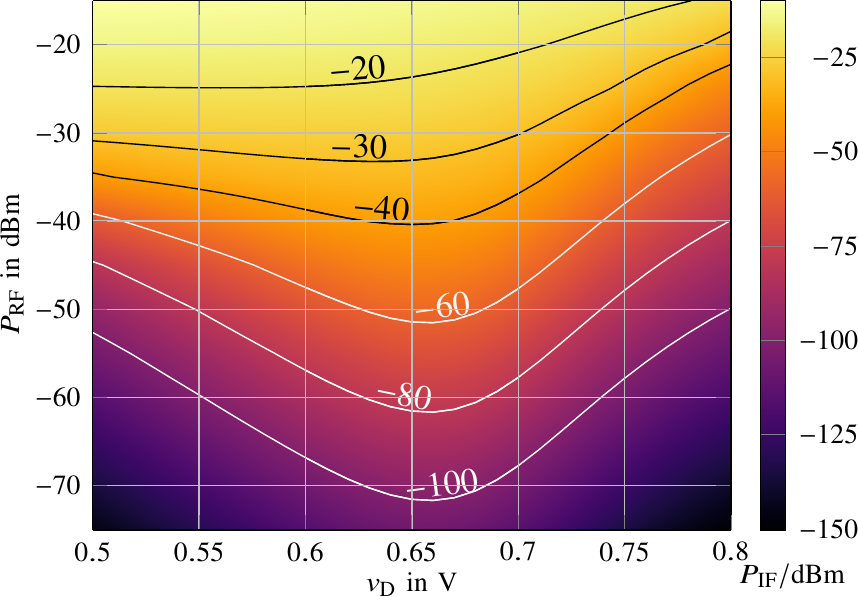}%
 \caption{Received IF power in the whole-Rx-circuit ADS simulation, considering a parameter sweep over biasing voltage and input power for a $37.5$/$38.5$\,GHz two-tone receive signal, whereby the weaker tone has the power $P_\mathrm{RF}-\SI{5}{dB}$.\label{fig:cont_bias}}
 \end{figure}

For the down-conversion, the Schottky diode MA4E1317 is employed~\cite{Macom}. 
First, the mixing behavior is analyzed in single-ended configuration. 
Thus, the $v$-$i$-characteristic and its first and second derivative are given in Fig.~\ref{fig:diode_dc}, according to an ADS DC simulation~\cite{ADS}. 
In this static operation point analysis, it seems that the best biasing point, i.e. the point with the highest-valued second order derivative, is achieved with an applied diode forward voltage $v_\mathrm D$ of about \SI{0.73}{\volt}. 
This is equivalent to a biasing current $i_\mathrm D$ of \SI{2.5}{\milli\ampere}. 
However, with the static DC analysis, the real world behavior at RF is not considered in a fully suitable manner. 
Many parasitic effects, like inductances and capacitances of the surrounding circuit as well as parasitic elements inside the diode, e.g. junction resistance and capacitance~\cite{Held}, will greatly influence the RF mixing behavior. 
Therefore, the whole feeding and filtering network has been simulated in ADS Momentum~\cite{ADS}. 
This includes the mentioned LNA with appropriately tapered connection lines for RF matching and a microstrip low pass filter before the diode to match the IF output signal, with a capacitance $C_\mathrm{LP}$ providing an IF connection to the following 50-$\Omega$ termination resistor $Z_0$ for the low-pass filter. 
Furthermore, two discrete inductors $L_\mathrm{bias}$ with \SI{8.2}{\nano\henry} from the L-05-series of Johanson Technologies~\cite{Johanson} in a 0201 package are employed to feed the diode by a current source. 
The inductors are placed after the microstrip low-pass filter and at the edge of the microstrip radial stub. Therefore, only IF signals are applied to them and their self-resonance frequency of 4.6\,GHz is sufficient. 
One of the inductors builds a bias-tee together with $C_\mathrm{LP}$ to prevent DC current flow into $Z_0$. 
A radial stub, $C_\mathrm{RS}$, after the diode helps to maximize the RF current flow through the diode~\cite{Chapman}. 
Hence, the strongest mixing products are produced with an RF short circuit following the diode in single ended configuration. 
Finally, a discrete output capacitor $C_\mathrm{IF}$ is used, offering a short circuit for the IF signals but blocking the DC signal. 
The discrete components as well as scattering parameter models of the LNA and the diode have been added to an ADS schematic and connected to the Momentum simulation. 
A harmonic balance analysis has been performed, applying a two-tone signal in the considered frequency range from 34\,GHz to 39\,GHz. 
The input power has been swept, as well as the biasing voltage of the diode. 
Fig.~\ref{fig:cont_bias} shows the simulation results for the 1-GHz IF mixing product power with the stated parameter sweep, for a 37.5/38.5\,GHz two-tone signal with tone powers $P_\mathrm{RF}$ and $P_\mathrm{RF}-\SI{5}{dB}$. 
For large input power, the operating point of the diode seems to be irrelevant due to the rectifying effect of the diode. 
For smaller input powers, the optimum bias voltage seems to lie around \SI{0.65}{\volt}. 
In Fig.~\ref{fig:cont_bias},  the frequency dependence of the mixer and amplifier is analyzed. 
A two-tone signal with 1\,GHz spacing and \SI{-40}{dBm} and \SI{-45}{dBm} input power is swept over the considered frequency range. 
Also for this analysis, it can be concluded that the optimum biasing point for the employed diode is at about \SI{0.65}{\volt}, resulting in a bias current of \SI{100}{\micro\ampere}, if all RF circuit and amplifier influences are taken into account.

\begin{figure}[t]
 \centering
    \includegraphics[]{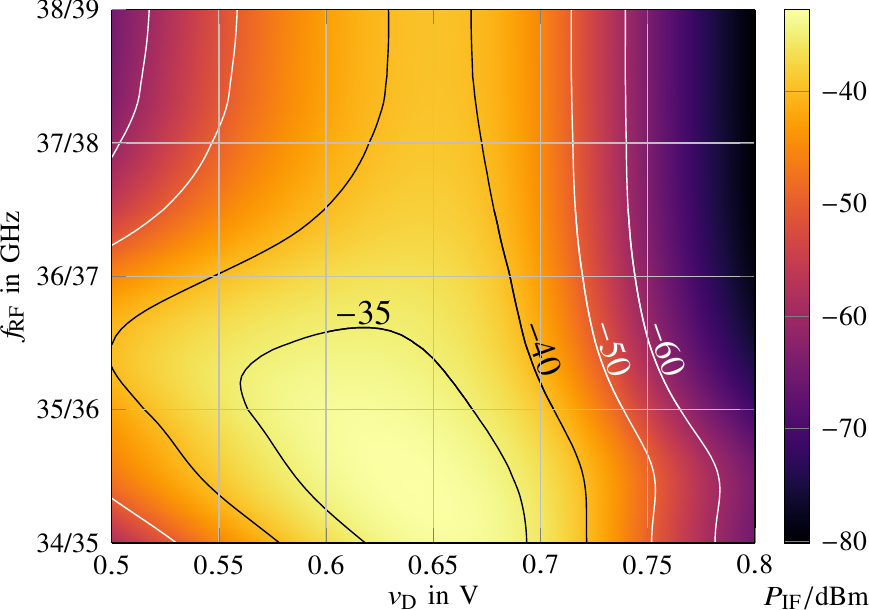}%
 \caption{Received IF power in the whole-Rx-circuit ADS simulation, considering a parameter sweep over biasing voltage and input frequency from a 34/35\,GHz two-tone signal to a 38/39\,GHz two-tone signal, whereby the tones have a power of \SI{-40}{dBm} and \SI{-45}{dBm}.\label{fig:cont_bias2}}
 \end{figure}%

As it can be recognized in Fig.~\ref{fig:cont_bias} and~\ref{fig:cont_bias2}, the mixing conversion loss is just about canceled by the gain of the LNA, if a two-tone input signal in the range of \SI{-40}{dBm} is assumed, since the IF signals show a very similar power level as applied at the input. 
For some frequencies and some input power levels, a positive overall conversion gain is realized. 

For the fabricated antennas, the LNA and diode are fed by a separate supply circuit for each antenna  from the rear side of the metal holding plate, including a low dropout voltage regulator for the drain voltage, a voltage inverter and a Zener diode for the gate voltage and an operational amplifier circuit to provide a current source for the diode biasing. 
Also, the coaxial cable connector is directed to the backside, since metal and dielectric objects around the antenna would disturb the radiation characteristics of the broad-beam antenna due to reflections and diffractions.

After the individual array elements, the commercial 8-to-1 power combiner ZN8PD1-53+ with about $0.5$\,dB loss at 1\,GHz and $0.9$\,dB loss at $2.5$\,GHz is employed~\cite{minicircuits}. 
Afterwards, the IF signal is amplified with about 20\,dB gain by a ZX60-V63+ wideband amplifier to improve the signal-to-noise ratio (SNR) at the measuring vector network analyzer (VNA)~\cite{minicircuits}. 
This makes sense since the coaxial cable attenuation amounts up to 10\,dB for the down-converted 2.5\,GHz IF signal. 
Hence, the VNA noise floor deteriorates with the IF amplifier  from $-100$ to $-90$\,dBm, with an employed  measurement bandwidth of~1\,kHz.

\section{Pattern Measurements} 
First, a reference measurement of a single self-mixing antenna element, illuminated by a horn antenna fed with a two-tone signal, was performed and compared to the multiplication of the radiation patterns obtained in a CST MWS simulation. 
Afterwards, the effect of the $4\times2$ self-mixing array was measured as compared to the single antenna element pattern. 
Both measurements verify the presented ideas. 

\subsection{The Pattern Measurement Setup} %
\begin{figure}[t]
 \centering
 \includegraphics[]{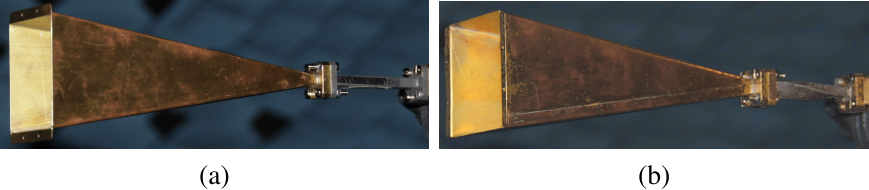}%
 \caption{The probe horn antenna in the both configurations. (a)~Straight feed, $E_\varphi$ polarization. (b)~\SI{90}{\degree} rotation, $E_\vartheta$ polarization.\label{fig:horn_antenna}}
\end{figure}

\begin{figure}[t]
 \centering
 \includegraphics[]{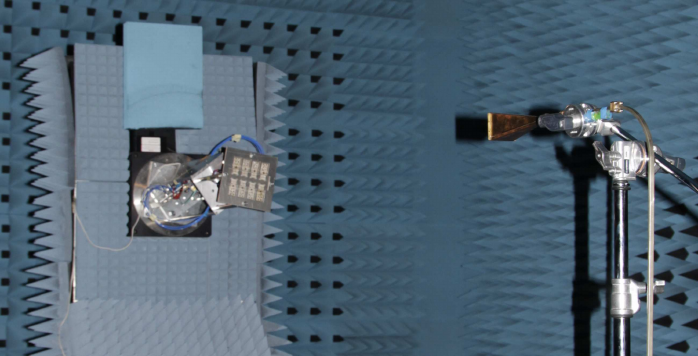}%
 \caption{AUT and probe horn.\label{fig:AUT_probe}}
\end{figure}

\begin{figure}[t]
 \centering
 \includegraphics[]{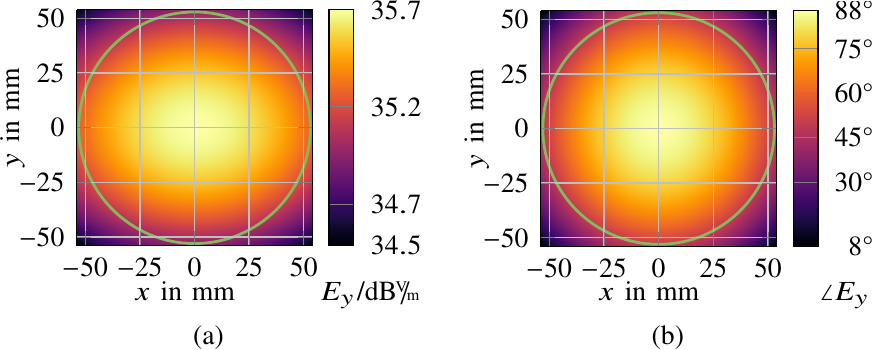}%
 \caption{AUT near-field illumination by the simulated horn antenna with an excitation power of 27\,dBm in \SI{1.5}{\meter} distance. (a)~Electric field magnitude at the AUT. (b)~Electric field phase at the AUT.\label{fig:horn_fields}}
\end{figure}

The self-mixing antenna is illuminated by a horn antenna fed with a two-tone signal for the measurements. 
After the down-conversion, the IF signal is measured with a VNA for different $\varphi$ and $\vartheta$ angles, whereby the AUT is rotated with a roll over azimuth positioner. 
A horn antenna fed by a $\mathrm{K_a}$-band hollow waveguide was employed as transmitting probe. 
The probe antenna rotation for measuring $E_\vartheta$ and $E_\varphi$ polarizations was achieved by a waveguide section which can be substituted by a twisted version of the same length. 
These two scenarios are depicted in Fig.~\ref{fig:horn_antenna}. 
The measurement scenario including the array AUT is depicted in Fig.~\ref{fig:AUT_probe}. 
The distance between AUT and probe is $1.5$\,m. 
I.e., the well-known far-field condition\footnote{The distance should be larger than $2D^2/\lambda$, where $D$ is the sum of the diameters of the minimum spheres around both antennas. $D$ has a value of about 105\,mm for the AUT plus 80\,mm for the probe horn, resulting in a far field distance of over 8\,m.} was not fulfilled. 
Therefore, the radiated field of the horn antenna was analyzed in a CST MWS simulation, with the results shown in Fig. \ref{fig:horn_fields}.  
The electric field magnitude is given in Fig. \ref{fig:horn_fields}(a) for an excitation power of 27\,dBm for $E_y$ polarization. 
The other field components have already decayed to values several orders of magnitude lower and, thus, can be neglected. 
The maximum area in which the AUT is rotated is given by the bright circle. 
It is observed that the probe gain varies by 0.5\,dB and the phase by less than  \SI{40}{\degree}. 
However, phase variations should be of little importance for the self-mixing receiver due to the already discussed effects.


For the feeding of the probe antenna, the signal generators Agilent E8247C and Rohde \& Schwarz SMR40 were employed. 
The transmit power was measured at the feeding point of the horn antenna to be 5\,dBm for the first generator and 0\,dBm for the second one. 
By doing so, transmission and combiner losses are taken into account. 
The horn antenna exhibits a gain of about 25\,dBi according to a CST MWS simulation. 
The Rx power is then calculated, assuming an Rx directivity of 0\,dBi in $\vartheta=\SI{0}{\degree}$ direction, according to the Friis equation as~\cite{Friis}
\begin{equation}
P_\mathrm{Rx}=P_\mathrm{Tx}+25\,\mathrm{dB}+\SI{20}{dB}\,\mathrm{lg}\frac{\lambda_\mathrm{RF}}{6\uppi\,\mathrm{m}}+\eta_\mathrm{tot}. 
\end{equation}
For the 0\,dBm input signal, $P_\mathrm{Rx,34\,GHz}=\SI{-43.4}{dBm}$ and $P_\mathrm{Rx,37.5\,GHz}=\SI{-44.3}{dBm}$. 
For the second signal generator with 5\,dBm transmit power, $P_\mathrm{Rx,36.5\,GHz}=\SI{-38.5}{dBm}$ and $P_\mathrm{Rx,38.5\,GHz}=\SI{-39.5}{dBm}$. 
The total efficiencies given in Fig.~\ref{fig:AUT_S11} are already taken into account in these numbers. 
The Rx circuit including LNA and mixing diode was simulated in ADS in a non-linear time domain analysis. 
For the 2.5\,GHz IF signal with input frequencies 34\,GHz and 36.5\,GHz, the expected output power is $-38$\,dBm. 
For the 1\,GHz signal with input frequencies 37.5\,GHz and 38.5\,GHz, the output power is a bit lower with $-41$\,dBm.
These values have been confirmed to an extent of some dB in the performed pattern measurements. 
The maximal observed received powers (in main beam direction) are $-18.2$\,dBm and $-20.9$\,dBm for the 1\,GHz and 2.5\,GHz IF signal, respectively, including about 20\,dB gain of the IF amplifier. 
It is observed that the approximate 3-dB difference is flipped in measurement due to fabrication tolerances, frequency-dependent behavior of all employed components and AUT gain fluctuations. 

\subsection{Single Self-Mixing Antenna Pattern Measurement} %
\begin{figure}[!p]
 \centering
 \includegraphics{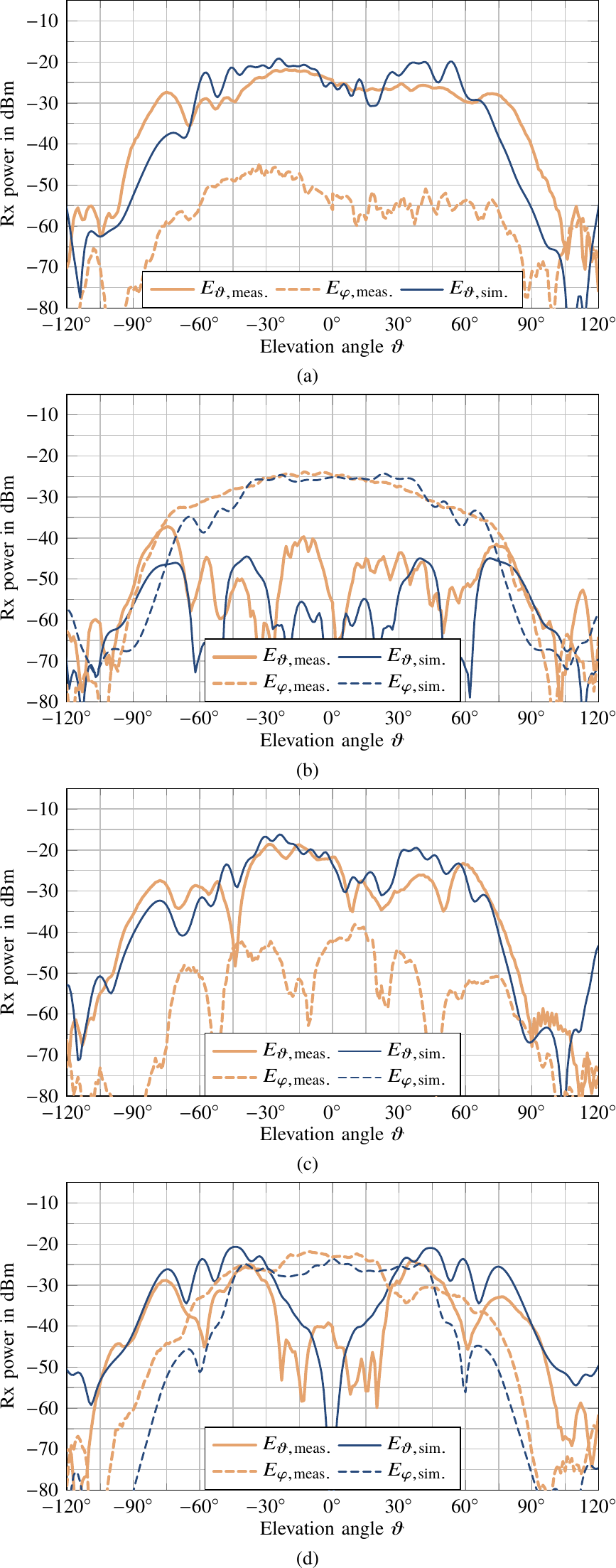}
 \caption{Single antenna self-mixing Rx pattern, in comparison of measurement and simulation. (a)~Rx pattern for a 34/36.5\,GHz two-tone signal, \textit{E}-plane. (b)~Rx pattern for a 34/36.5\,GHz two-tone signal, \textit{H}-plane. (c)~Rx pattern for a 37.5/38.5\,GHz two-tone signal, \textit{E}-plane. (d)~Rx pattern for a 37.5/38.5\,GHz two-tone signal, \textit{H}-plane.\label{fig:single_pattern}}
\end{figure}

According to~\eqref{eq:pattern}, the Rx pattern of a self-mixing antenna is represented by the multiplication of the two radiation patterns of the employed antenna at the two frequencies of the two-tone signal, if an ideal square-law detector is assumed. 
Therefore, the reference radiation patterns were attained in a CST MWS simulation, taking into account the real holding structures, i.e. the metal plate below and surrounding the antenna as well as the metal and polyimide screws fixing the substrate. 
The ripple present in measurement and simulation originates from substrate wave radiation and metal plate reflection. 
In Fig.~\ref{fig:single_pattern}, the measurement results for the two polarizations $E_\vartheta$ and $E_\varphi$ are compared to the simulation results (multiplication of the two simulated radiation patterns), scaled to the same down-converted Rx power as attained in the measurement. 
The cross-polarization $E_{\vartheta,\mathrm{sim}}$ obtained from simulation has very low values and does not appear in Fig.~\ref{fig:single_pattern}(a). 
The \textit{E}- and \textit{H}-planes are defined for the rectangular patch mode, as the $\varphi\!=\!\SI{90}{\degree}$- and  $\varphi\!=\!\SI{0}{\degree}$-planes. 
Measurement and simulation agree quite well, in particular if it is taken into account that simulation inaccuracies are doubled by the pattern multiplication.

\begin{figure}[!p]
 \centering
 \includegraphics{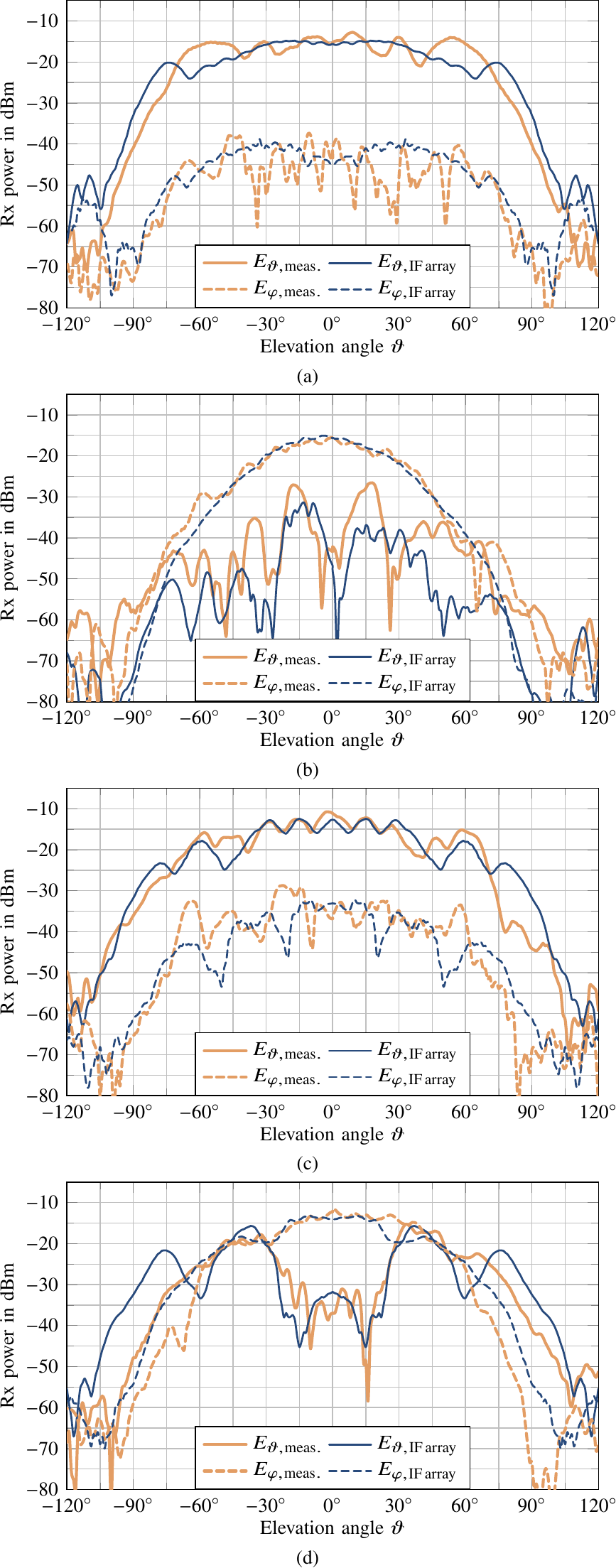}
 \caption{Measured self-mixing array Rx pattern, as compared to the measured single element pattern times the ideal IF array factors. (a)~Rx pattern for a 34/36.5\,GHz two-tone signal, \textit{E}-plane. (b)~Rx pattern for a 34/36.5\,GHz two-tone signal, \textit{H}-plane. (c)~Rx pattern for a 37.5/38.5\,GHz two-tone signal, \textit{E}-plane. (d)~Rx pattern for a 37.5/38.5\,GHz two-tone signal, \textit{H}-plane.\label{fig:array_pattern}}
\end{figure}

\subsection{Self-Mixing Array Pattern Measurement} %
To evaluate the self-mixing array effects, a $4\times2$ array of the presented single antenna element  was fabricated and measured in the exactly same setup, with the same incident power and probe setup. 
The antenna signals were added with a coherent eight-to-one combiner. 
Therefore, 9\,dB more receive power is expected with 8 array elements.  
The center-to-center distance of the array elements  was about 32\,mm in $x$-direction and 36\,mm in $y$-direction, according to the coordinate system in Fig.~\ref{fig:AUT_circuit}.  
For the receiving frequency range, this is equal to about $4\lambda_\mathrm{RF}$ and $4.5\lambda_\mathrm{RF}$, with $\lambda_\mathrm{RF}\approx\SI{8}{\milli\meter}$ being the free-space wavelength in the considered frequency range (at 36\,GHz). 
However, according to the self-mixing array principles explained in Section II, only the phase-shift at the difference frequency of  the two-tone signal is important for the eight-to-one power combiner. 
Hence, the array factor according to~\eqref{eq:AF} is evaluated at 1\,GHz and 2.5\,GHz, for the two different receive signals and in the $x=0$ and $y=0$ cutting planes. 
As a result, the effective element spacing is reduced from about several RF wavelengths to only $0.1\lambda_{\mathrm{RF}}$ and  $0.15\lambda_{\mathrm{RF}}$ for 1\,GHz and $0.25\lambda_{\mathrm{RF}}$ and  $0.275\lambda_{\mathrm{RF}}$ for 2.5\,GHz IF frequency. 
This results in a very wide array factor while maintaining the large array gain for 8 elements as it is shown in Fig. \ref{fig:array_pattern}, where the measured array is compared to the measured single element characteristic times the ideally expected array factor. 
Due to the lack of symmetry in the single antenna measurement in the \textit{E}-plane, the \textit{E}-plane receive characteristics of the single antenna are summed up including a mirrored version representing the row of rotated antennas. 
Thus, a symmetric single element pattern is utilized  for the comparison with the array. 
This is reasonable since the array mounting of the rotated antennas is somehow symmetrical. 
The agreement between single element measured Rx characteristic and measured self-mixing array Rx pattern is very good. 
In the \textit{H}-plane, a weak beamforming effect is observed due to the  number of antennas. 
The array gain in $\vartheta\!=\!\SI{0}{\degree}$ direction is about 9\,dB, just as expected.

\begin{figure}[t]
\centering
 \includegraphics{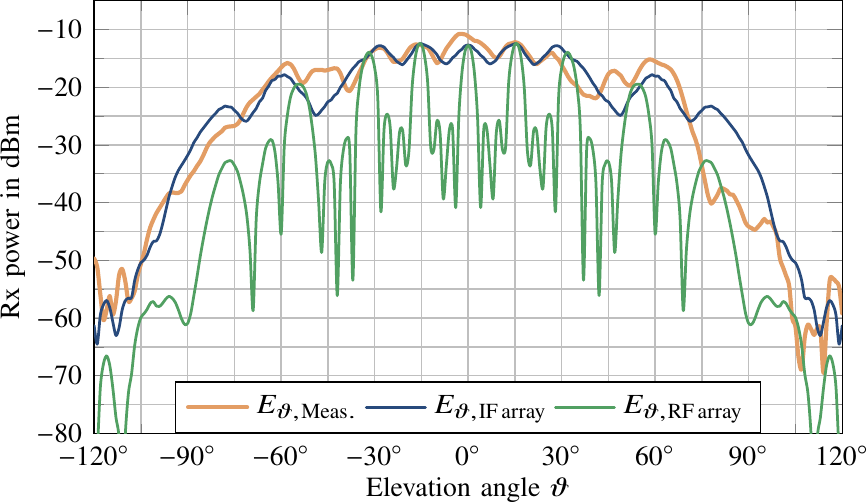}%
\caption{Measured Self-mixing Array Rx pattern in the \textit{E}-plane, as compared to the measured single element pattern times the ideal array factors at IF (1\,GHz) and RF (38.5\,GHz).\label{fig:array_pattern_RF}}
\vspace*{-0.35cm}
\end{figure}

Two interesting facts can be stated: 
First, the two antenna rows of the array have a phase shift of 180\si{\degree} with respect to each other, due to the row rotation of the second antennas. 
This is completely compensated by the self-mixing. 
In a convential array with a coherent equal-phase power combiner, the antennas would produce destructive interference in the \mbox{\textit{xz}-plane} leading to absolutely no received power in the \mbox{\textit{H}-plane}. 
Second, the array can be compared to a sparse array with convential RF beamforming. 
For this case, the rotation of the second row of antennas is neglected for logical reasons. 
In this comparison in Fig.~\ref{fig:array_pattern_RF}, significant grating lobes are observed in the RF pattern.
The great advantage of the broad Rx beamwidth of self-mixing antenna arrays is clearly visible. 

\section{Conclusion} 
The principle of self-mixing antenna arrays was discussed and a new formulation for the array factor at an intermediate frequency (IF), for the self-mixing of a two-tone signal, has been derived. 
Afterwards, a receive circuit was designed and optimized for the application in self-mixing arrays for wireless communications. 
The employed antenna shows a large bandwidth and a broad beamwidth. 
The receive circuit contains a low noise amplifier to amplify the receive signal, and a subsequent Schottky diode with approximately quadratic voltage-current characteristic for down-conversion, biased for the optimal operating point. 
Finally, measurements were carried out for IF patterns with different radio frequency (RF) transmit signals. 
It was shown that the single self-mixing antenna works as expected, multiplying the radiation patterns at the different RF frequencies. 
Furthermore, a $4\times2$ self-mixing array was set up. 
Its pattern measurements show very good agreement with the multiplication of measured single element pattern times the ideal array factor, proving the presented theoretical derivation of the self-mixing IF array factor.

\bibliographystyle{IEEEtran}

\bibliography{IEEEabrv,ref}

\vfill
\newpage

\begin{IEEEbiography}
  [{\includegraphics[width=1in,height=1.25in,clip,keepaspectratio,]{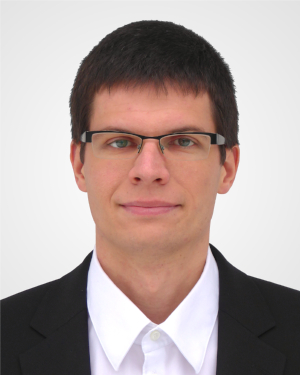}}]
  {Jonas Kornprobst} (S'17) received the B.Eng.\ degree from the University of 
  Applied Sciences Rosenheim, Rosenheim, Germany, and the M.Sc.\ degree from the 
  Technical University of Munich, Munich, Germany, in 2014 and 2016, 
  respectively, both in electrical engineering and information technology. Since 
  April 2016, he has been a Research Assistant at the Chair of High-Frequency 
  Engineering, Technical University of Munich, Munich, Germany. His current 
  research interests include numerical electromagnetics, in particular integral 
  equation methods, antenna and antenna array design as well as microwave circuits.
\end{IEEEbiography}
\begin{IEEEbiography}
  [{\includegraphics[width=1in,height=1.25in,clip,keepaspectratio]{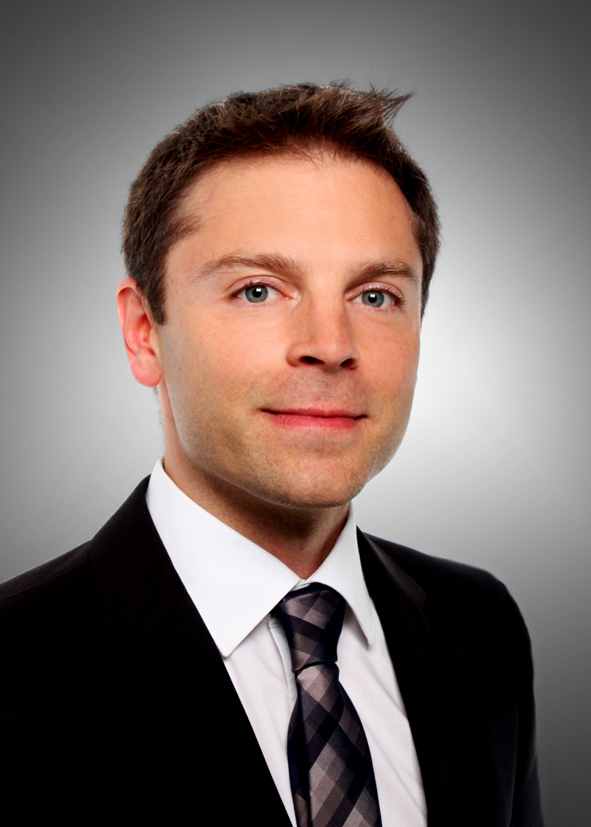}}]
{Thomas J. Mittermaier} (S'12$-$M'17) received the Dipl.-Ing.(FH)
degree from Fachhochschule Regensburg, Regensburg, Germany, and
the M.Sc. degree from the Technical University of Munich, Munich,
Germany, in 2009 and 2011, respectively, both in electrical
engineering and information technology. From 2011 to 2017 he was
a Research Assistant at the Chair of High-Frequency Engineering of
the Technical University of Munich, Munich, Germany. His doctoral
research focused on signal processing and tracking algorithms for
extremely short-range radar in industrial machine tools as well as
microwave circuits.
In May 2017, he joined the RF360 Europe GmbH, Munich, as a Development
Engineer for bulk acoustic wave devices. He is currently involved in
linear and nonlinear modeling, in power durability and life-time testing
of acoustic resonators and filters for future mobile communications.
\end{IEEEbiography}

\begin{IEEEbiography}
  [{\includegraphics[width=1in,height=1.25in,clip,keepaspectratio]{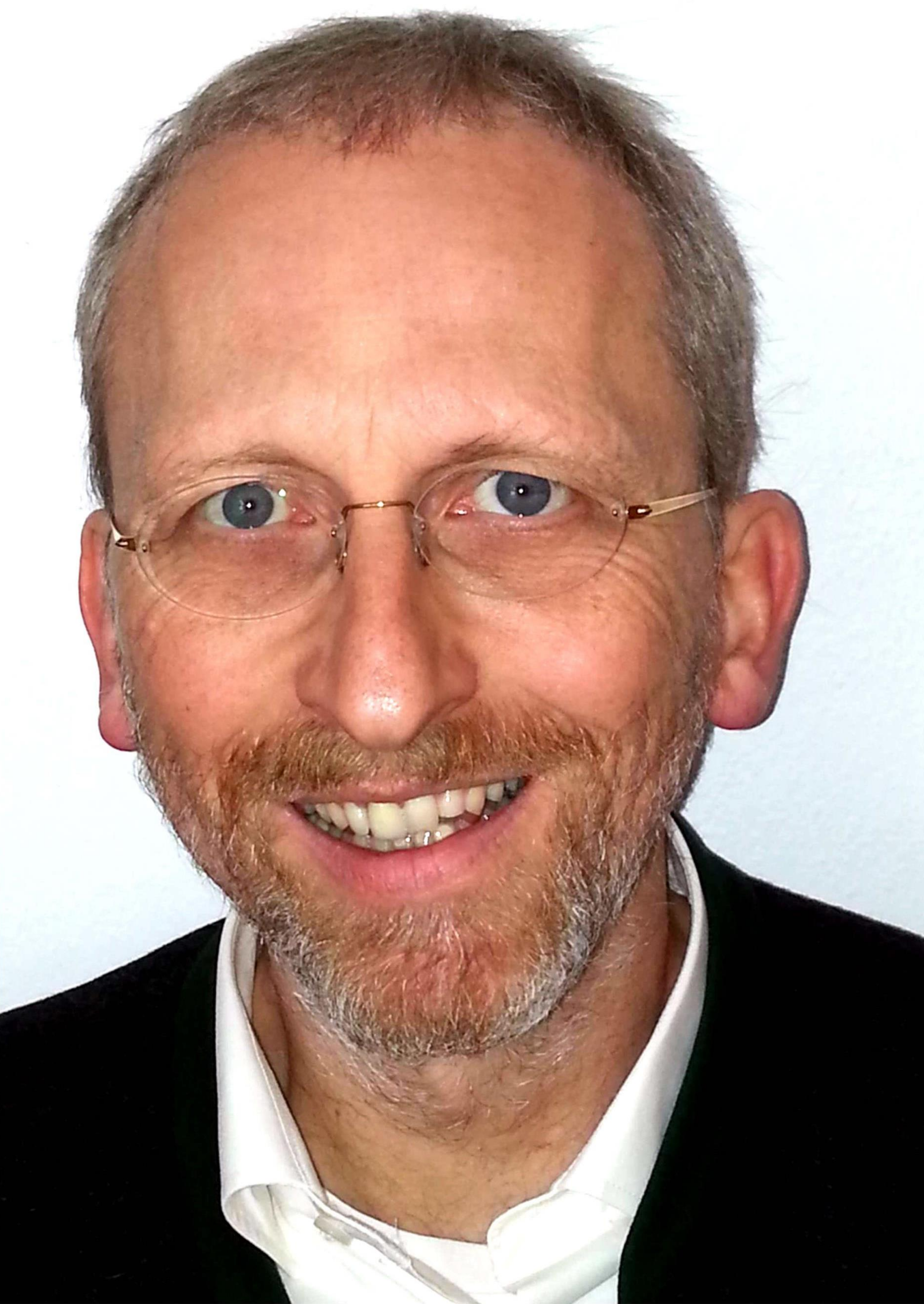}}]
  {Thomas F. Eibert} (S'93$-$M'97$-$SM'09) received the Dipl.-Ing.(FH) degree
  from Fachhochschule N\"urnberg, Nuremberg, Germany, the Dipl.-Ing.\ degree from
  Ruhr-Universit\"at Bochum, Bochum, Germany, and the Dr.-Ing.\ degree from
  Bergische Universit\"at Wuppertal, Wuppertal, Germany, in 1989, 1992, and
  1997, respectively, all in electrical engineering. From 1997 to 1998, he was
  with the Radiation Laboratory, EECS Department at the University of Michigan,
  Ann Arbor, MI, USA, from 1998 to 2002, he was with Deutsche Telekom,
  Darmstadt, Germany, and from 2002 to 2005, he was with the Institute for
  High-Frequency Physics and Radar Techniques of FGAN e.V., Wachtberg, Germany,
  where he was Head of the Department Antennas and Scattering. From 2005 to
  2008, he was a Professor of radio frequency technology at Universit\"at
  Stuttgart, Stuttgart, Germany. Since October 2008, he has been a Professor of
  high-frequency engineering at the Technical University of Munich, Munich,
  Germany. His major areas of interest are numerical electromagnetics, wave
  propagation, measurement and field transformation techniques for antennas and
  scattering as well as all kinds of antenna and microwave circuit technologies
  for sensors and communications.
\end{IEEEbiography}

\vfill

\end{document}